\documentclass[twocolumn,preprintnumbers,superscriptaddress,showpacs,amsmath,amssymb]{revtex4}

\usepackage[dvips]{epsfig,color}
\include{graphics}

\usepackage{graphicx}
\usepackage{dcolumn}

\begin{document}

\newcommand*{\NOWSCAROLINA}{University of South Carolina, Columbia, SC 29208.}
\newcommand*{\NOWMC}{Montgomery College, Rockville, MD 20850.}


\newcommand*{\GWU}{The George Washington University, Washington, DC 20052}
\newcommand*{\GWUindex}{14}
\affiliation{\GWU}
\newcommand*{\JLAB}{Thomas Jefferson National Accelerator Facility, Newport News, Virginia 23606}
\newcommand*{\JLABindex}{33}
\affiliation{\JLAB}
\newcommand*{\ANL}{Argonne National Laboratory}
\newcommand*{\ANLindex}{1}
\affiliation{\ANL}
\newcommand*{\ASU}{Arizona State University, Tempe, Arizona 85287-1504}
\newcommand*{\ASUindex}{2}
\affiliation{\ASU}
\newcommand*{\UCLA}{University of California at Los Angeles, Los Angeles, California  90095-1547}
\newcommand*{\UCLAindex}{3}
\affiliation{\UCLA}
\newcommand*{\CSU}{California State University, Dominguez Hills, Carson, CA 90747}
\newcommand*{\CSUindex}{4}
\affiliation{\CSU}
\newcommand*{\CMU}{Carnegie Mellon University, Pittsburgh, Pennsylvania 15213}
\newcommand*{\CMUindex}{5}
\affiliation{\CMU}
\newcommand*{\CUA}{Catholic University of America, Washington, D.C. 20064}
\newcommand*{\CUAindex}{6}
\affiliation{\CUA}
\newcommand*{\SACLAY}{CEA, Centre de Saclay, Irfu/Service de Physique Nucl\'eaire, 91191 Gif-sur-Yvette, France}
\newcommand*{\SACLAYindex}{7}
\affiliation{\SACLAY}
\newcommand*{\CNU}{Christopher Newport University, Newport News, Virginia 23606}
\newcommand*{\CNUindex}{8}
\affiliation{\CNU}
\newcommand*{\UCONN}{University of Connecticut, Storrs, Connecticut 06269}
\newcommand*{\UCONNindex}{9}
\affiliation{\UCONN}
\newcommand*{\ECOSSEE}{Edinburgh University, Edinburgh EH9 3JZ, United Kingdom}
\newcommand*{\ECOSSEEindex}{10}
\affiliation{\ECOSSEE}
\newcommand*{\FU}{Fairfield University, Fairfield CT 06824}
\newcommand*{\FUindex}{11}
\affiliation{\FU}
\newcommand*{\FIU}{Florida International University, Miami, Florida 33199}
\newcommand*{\FIUindex}{12}
\affiliation{\FIU}
\newcommand*{\FSU}{Florida State University, Tallahassee, Florida 32306}
\newcommand*{\FSUindex}{13}
\affiliation{\FSU}
\newcommand*{\ECOSSEG}{University of Glasgow, Glasgow G12 8QQ, United Kingdom}
\newcommand*{\ECOSSEGindex}{15}
\affiliation{\ECOSSEG}
\newcommand*{\ISU}{Idaho State University, Pocatello, Idaho 83209}
\newcommand*{\ISUindex}{16}
\affiliation{\ISU}
\newcommand*{\INFNFR}{INFN, Laboratori Nazionali di Frascati, 00044 Frascati, Italy}
\newcommand*{\INFNFRindex}{17}
\affiliation{\INFNFR}
\newcommand*{\INFNGE}{INFN, Sezione di Genova, 16146 Genova, Italy}
\newcommand*{\INFNGEindex}{18}
\affiliation{\INFNGE}
\newcommand*{\INFNRO}{INFN, Sezione di Roma Tor Vergata, 00133 Rome, Italy}
\newcommand*{\INFNROindex}{19}
\affiliation{\INFNRO}
\newcommand*{\ORSAY}{Institut de Physique Nucl\'eaire ORSAY, Orsay, France}
\newcommand*{\ORSAYindex}{20}
\affiliation{\ORSAY}
\newcommand*{\ITEP}{Institute of Theoretical and Experimental Physics, Moscow, 117259, Russia}
\newcommand*{\ITEPindex}{21}
\affiliation{\ITEP}
\newcommand*{\JMU}{James Madison University, Harrisonburg, Virginia 22807}
\newcommand*{\JMUindex}{22}
\affiliation{\JMU}
\newcommand*{\KYUNGPOOK}{Kyungpook National University, Daegu 702-701, Republic of Korea}
\newcommand*{\KYUNGPOOKindex}{23}
\affiliation{\KYUNGPOOK}
\newcommand*{\MIT}{Massachusetts Institute of Technology, Cambridge, Massachusetts  02139-4307}
\newcommand*{\MITindex}{24}
\affiliation{\MIT}
\newcommand*{\NSU}{Norfolk State University, Norfolk, Virginia 23504}
\newcommand*{\NSUindex}{25}
\affiliation{\NSU}
\newcommand*{\OHIOU}{Ohio University, Athens, Ohio  45701}
\newcommand*{\OHIOUindex}{26}
\affiliation{\OHIOU}
\newcommand*{\ODU}{Old Dominion University, Norfolk, Virginia 23529}
\newcommand*{\ODUindex}{27}
\affiliation{\ODU}
\newcommand*{\RPI}{Rensselaer Polytechnic Institute, Troy, New York 12180-3590}
\newcommand*{\RPIindex}{28}
\affiliation{\RPI}
\newcommand*{\URICH}{University of Richmond, Richmond, Virginia 23173}
\newcommand*{\URICHindex}{29}
\affiliation{\URICH}
\newcommand*{\ROMAII}{Universita' di Roma Tor Vergata, 00133 Rome Italy}
\newcommand*{\ROMAIIindex}{30}
\affiliation{\ROMAII}
\newcommand*{\MOSCOW}{Skobeltsyn Nuclear Physics Institute, Skobeltsyn Nuclear Physics Institute, 119899 Moscow, Russia}
\newcommand*{\MOSCOWindex}{31}
\affiliation{\MOSCOW}
\newcommand*{\SCAROLINA}{University of South Carolina, Columbia, South Carolina 29208}
\newcommand*{\SCAROLINAindex}{32}
\affiliation{\SCAROLINA}
\newcommand*{\UNIONC}{Union College, Schenectady, NY 12308}
\newcommand*{\UNIONCindex}{34}
\affiliation{\UNIONC}
\newcommand*{\UTFSM}{Universidad T\'{e}cnica Federico Santa Mar\'{i}a, Casilla 110-V Valpara\'{i}so, Chile}
\newcommand*{\UTFSMindex}{35}
\affiliation{\UTFSM}
\newcommand*{\VIRGINIA}{University of Virginia, Charlottesville, Virginia 22901}
\newcommand*{\VIRGINIAindex}{36}
\affiliation{\VIRGINIA}
\newcommand*{\WM}{College of William and Mary, Williamsburg, Virginia 23187-8795}
\newcommand*{\WMindex}{37}
\affiliation{\WM}
\newcommand*{\YEREVAN}{Yerevan Physics Institute, 375036 Yerevan, Armenia}
\newcommand*{\YEREVANindex}{38}
\affiliation{\YEREVAN}

\newcommand*{\NOWUK}{University of Kentucky, LEXINGTON, KENTUCKY 40506}
\newcommand*{\NOWJLAB}{Thomas Jefferson National Accelerator Facility, Newport News, Virginia 23606}
\newcommand*{\NOWLANL}{Los Alamos National Laborotory, New Mexico, NM}
\newcommand*{\NOWCNU}{Christopher Newport University, Newport News, Virginia 23606}
\newcommand*{\NOWECOSSEE}{Edinburgh University, Edinburgh EH9 3JZ, United Kingdom}
\newcommand*{\NOWWM}{College of William and Mary, Williamsburg, Virginia 23187-8795}
\author {R.~Nasseripour} \affiliation{\GWU}
\author {B.L.~Berman} \affiliation{\GWU}
\author {N.~Benmouna} 
\altaffiliation[Current address:]{\NOWMC}
\affiliation{\GWU}
\author {Y.~Ilieva} 
\altaffiliation[Current address:]{\NOWSCAROLINA}
\affiliation{\GWU} 
\author {J.M.~Laget} 
\affiliation{\JLAB}
\author {K. P. ~Adhikari} 
\affiliation{\ODU}
\author {M.J.~Amaryan} 
\affiliation{\ODU}
\author {M.~Anghinolfi}
\affiliation{\INFNGE}
\author {H.~Baghdasaryan} 
\affiliation{\VIRGINIA}
\author {J.~Ball} 
\affiliation{\SACLAY}
% ...

\collaboration{The CLAS Collaboration}
\noaffiliation


%
 
%
%
%
%
%
%

\title{Photodisintegration of $^4$He into p+t}
 
\date{\today}

\begin{abstract}

The two-body photodisintegration of $^4$He into a proton and a triton has been 
studied using the CEBAF Large-Acceptance Spectrometer (CLAS) at Jefferson Laboratory. 
Real photons produced with the Hall-B bremsstrahlung-tagging system in the energy 
range from 0.35 to 1.55 GeV were incident on a liquid $^4$He target. This is the 
first measurement of the photodisintegration of $^4$He above 0.4 GeV. The differential 
cross sections for the $\gamma$$^4$He$\rightarrow pt$ reaction have been measured as 
a function of photon-beam energy and proton-scattering angle, and are compared with 
the latest model calculations by J.-M. Laget. At 0.6-1.2 GeV, our data are in good 
agreement only with the calculations that include three-body mechanisms, thus 
confirming their importance. 
These results reinforce the conclusion of our previous study of the 
three-body breakup of $^3$He that demonstrated the great importance of three-body 
mechanisms in the energy region 0.5-0.8 GeV . 
\end{abstract}


\maketitle


\section{Introduction}
\label{sec:intro}

One of the difficult challenges of nuclear physics is to understand
the nature of the strong many-body interaction among the
nucleons in the nucleus. In particular, understanding the contribution and  
manifestations of three-body forces is an important ingredient of the 
theoretical calculations that attempt to describe the reaction mechanisms.     
Photonuclear reactions are induced by a well known probe, and are 
especially sensitive 
to meson-exchange currents and isobar degrees of freedom. 

The two-body photodisintegration of $^4$He into a proton and a triton has been
studied over the years in the low- and intermediate-energy regions, up to 0.4 GeV, 
where the one- and two-body mechanisms dominate the reaction 
\cite{kiergan,argan,arends,schumacher,jones}. 
The higher photon-energy region used in this experiment allows us to access
larger momentum transfers, where the three-body mechanisms, mostly through
higher-mass-meson double scattering, are expected to make a larger contribution.

\begin{figure}[htbp]
 \begin{center}
 \mbox{\epsfxsize=4.0cm\leavevmode \epsffile{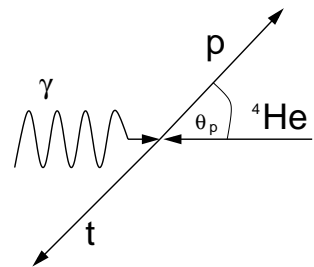}}	
 \end{center}
\caption{\small{Diagram of the $\gamma ^4$He $\rightarrow pt$ reaction 
in the center-of-mass frame.
The angle between the beam direction and the emitted proton in the center-of-mass
frame is denoted as $\theta_p$.}}
\label{fig:diagram}
\end{figure}

In this analysis,
the differential cross sections for the $\gamma ^4$He $\rightarrow pt$ reaction
were measured as a function of 
photon energy from 0.35 to 1.55 GeV, and over a wide range of the proton-scattering angle 
in the center-of-mass frame, as shown in Fig. \ref{fig:diagram}. 
These measurements are complementary to the 
three-body breakup of $^3$He for the study of three-body
reaction mechanisms \cite{niccolai}. 
The results are compared with the latest model predictions
of J.-M. Laget \cite{laget,laget2}, where three-body mechanisms for this
channel are included in the calculations.

\section{Model Predictions}
\label{sec:theory}

\begin{figure}[htbp]
 \begin{center}
 \mbox{\epsfxsize=6.5cm\leavevmode \epsffile{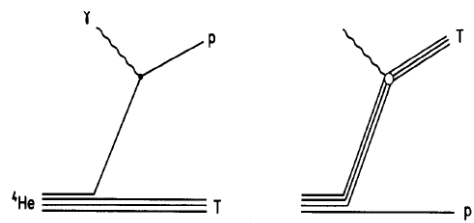}}
 \end{center}
\caption{\small{The one-body mechanisms included in the Laget
model showing proton (left) and triton (right) exchange. This figure is
from Ref. \cite{laget}.}}
\label{fig:onebody}
\end{figure}
\begin{figure}[htbp]
 \begin{center}
 \mbox{\epsfxsize=7.5cm\leavevmode \epsffile{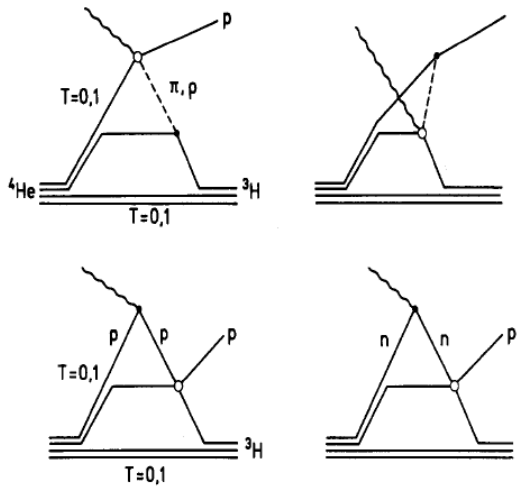}}	
 \end{center}
\caption{\small{The two-body mechanisms. The top diagrams show $\pi$ and $\rho$ 
exchange. The bottom diagrams show the nucleon-nucleon rescattering final-state interactions.
The diagrams on the right come from the antisymmetry of the two active nucleons.
This figure is from Ref. \cite{laget}.}}
\label{fig:twobody}
\end{figure}
\begin{figure}[htbp]
 \begin{center}
 \mbox{\epsfxsize=9.0cm\leavevmode \epsffile{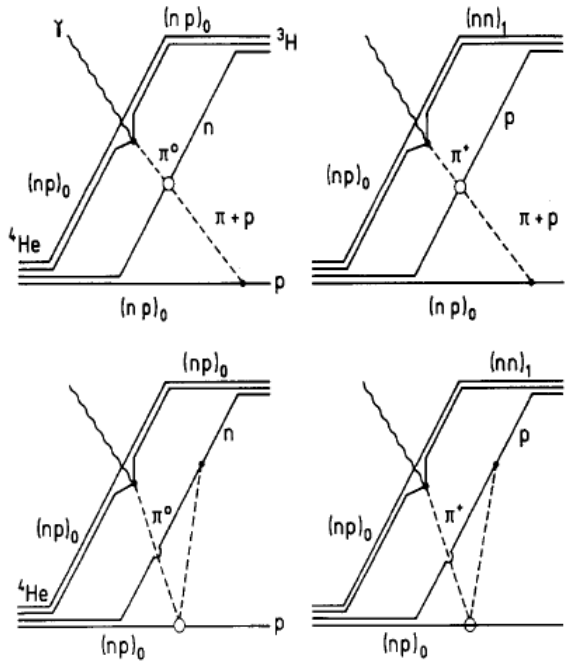}}	
 \end{center}
\caption{\small{The three-body mechanisms showing meson double scattering.
The diagrams on the bottom correspond to the antisymmetry of the two nucleons in 
the pair that absorbs the intermediate pion.
This figure is from Ref. \cite{laget}.}}
\label{fig:threebody}
\end{figure}

A calculation has been performed by J.-M. Laget \cite{laget} in which
three types of reaction mechanisms for this channel have been included. 
Figures \ref{fig:onebody}, \ref{fig:twobody}, and \ref{fig:threebody} show the diagrams 
for one-body, two-body, and three-body reaction mechanisms, respectively. 
The one-body mechanisms include the proton and triton (three-nucleon) exchange diagrams 
where a proton or a triton knockout is assumed. 
The proton-exchange diagram is dominant when the proton is emitted at forward angles. 
The two-body mechanisms included in this calculation are the two-nucleon meson exchange
(including the pion and the rho meson)  
and nucleon-nucleon final-state interactions. The diagrams due to the 
antisymmetry of the two active nucleons are also included in the two-body mechanisms. 
The three-body mechanisms include the 
meson double-scattering amplitudes. As shown in the bottom panels of 
Fig. \ref{fig:threebody}, diagrams that correspond to the antisymmetry of 
the two nucleons in the pair that absorbs the pion are also included in the model.

\begin{figure}[htbp]
 \begin{center}
 \mbox{\epsfxsize=8.0cm\leavevmode \epsffile{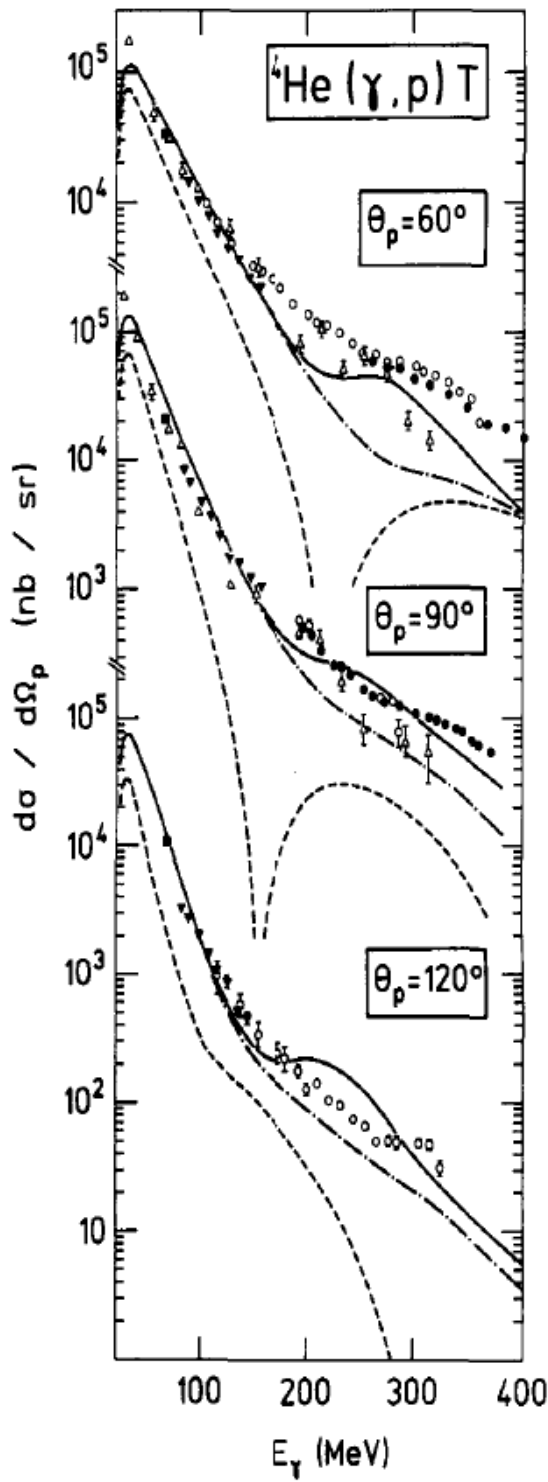}}	
 \end{center}
\caption{\small{Previous world data for the photodisintegration of $^4$He into $p + t$, 
plotted as a function of the incident photon energy (open triangles \cite{kiergan}, 
closed squares \cite{argan}, closed circles \cite{arends},
open circles \cite{schumacher}, and closed diamonds \cite{jones}) compared with Laget's
model calculations, including one-body diagrams only (dashed lines), two-body diagrams
(dot-dashed lines), and three-body diagrams (solid lines). This figure is from Ref. \cite{laget}.}}
\label{fig:laget}
\end{figure}
The two- and three-body diagrams included in this model 
were found to be dominant in the study of two- and three-body 
disintegration of $^3$He. The relevance of this description is  
investigated here in a different environment, the $^4$He nucleus, having markedly
different n-body density and wave function.  

Results of the model calculations are compared with previous world data in Fig. \ref{fig:laget},
where the cross sections are plotted as a function of the incident photon energy for 
three values of the proton angle with the incident 
photon-beam direction, $\theta_p = 60^{\circ}$, $\theta_p = 90^{\circ}$, and $\theta_p = 120^{\circ}$.
The comparison with Laget's model \cite{laget} reveals that three-body mechanisms must be
taken into account to describe the data. 

This model has been extended to higher energies according to Ref. \cite{laget2}
and improved using  i) the full relativistic expression for the 
nucleon currents, ii) a high-energy diffractive NN scattering amplitude,
and iii) the latest ground-state wave function \cite{forest} that has
been generated from the Argonne AV18 NN potential \cite{av18} and 
the Urbana UIX three-nucleon force \cite{UR-IX}.

A thorough investigation of three-body mechanisms was performed earlier, in the analysis of the 
reaction $\gamma ^3$He $\rightarrow ppn$ from the CLAS g3a data, which
revealed that three-body mechanisms are most prominent in the energy range from 
0.5 to 0.8 GeV \cite{niccolai}. Three-body mechanisms also have been studied
in the two-body photodisintegration of $^3$He \cite{laget3}.
The results presented here are the first to include
the $\gamma ^4$He $\rightarrow pt$ reaction at energies higher than 0.4 GeV.


\section{Experiment and Data Analysis}
\label{sec:exper}

\subsection{Experimental Apparatus}

\begin{figure}[htbp]
 \begin{center}
 \mbox{\epsfxsize=8.0cm\leavevmode \epsffile{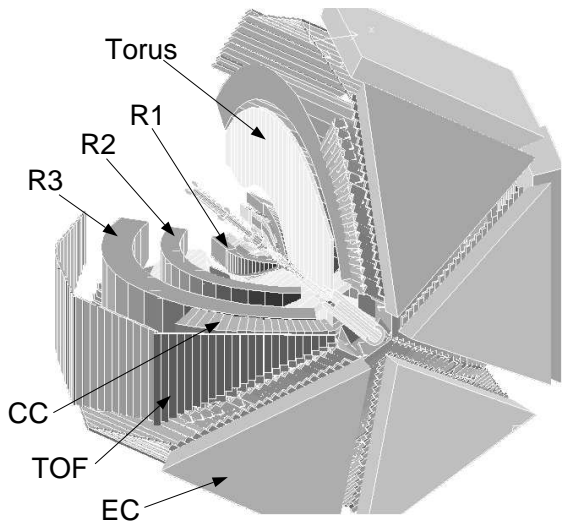}}	
 \mbox{\epsfxsize=9.5cm\leavevmode \epsffile{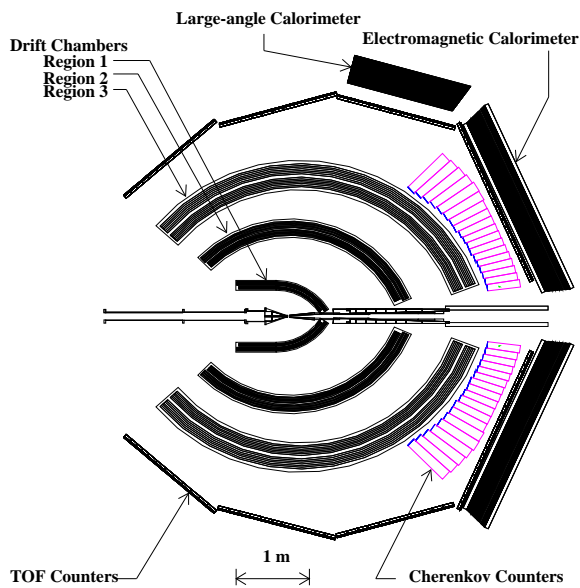}}
 \end{center}
\caption{\small{CLAS is a nearly 4$\pi$-sr detector system used to observe
multiparticle final states. Top: Three-dimensional representation 
of CLAS, with a portion of the system cut away to highlight  
elements of the detector system as described in the text. Bottom: Midplane slice of the CLAS.}}
\label{fig:clas1}
\end{figure}

The $\gamma ^4$He $\rightarrow pt$ reaction was measured during the g3a experiment in December 1999
with the CEBAF Large-Acceptance Spectrometer (CLAS) \cite{mecking} at Jefferson Lab, shown in Fig. \ref{fig:clas1}.
CLAS is a large acceptance spectrometer used to detect multiparticle final states.
Six superconducting coils generate a toroidal magnetic field around the target with
azimuthal symmetry about the beam axis. The coils divide CLAS into six sectors,
each functioning as an independent magnetic spectrometer. Each sector is instrumented
with three regions of drift chambers (DCs), R1-3, to determine charged-particle trajectories \cite{dc},
scintillator counters (SCs) for time-of-flight measurements \cite{sc}, and, 
in the forward region, gas-filled threshold Cherenkov counters (CCs) for electron/pion
separation up to 2.5 GeV \cite{cc}, and electromagnetic calorimeters (ECs) to identify
and measure the energy of electrons and high-energy neutral particles, as well as 
to provide electron/pion separation above 2.5 GeV \cite{ec}.
In the g3a experiment, real photons produced with the Jefferson Lab Hall-B bremsstrahlung-tagging 
system \cite{sober} in the energy range from 0.35 to 1.55 GeV were incident on 
an 18-cm-thick liquid $^4$He target.

The photon beam was produced via bremsstrahlung from the primary electron beam operating at 1.645 GeV. Electrons were incident on the 
thin radiator of the Hall-B Photon Tagger \cite{sober}. 
Tagged photons were produced with 20-95\%
of the energy of the primary electron beam.
About 10$^9$ triggers were collected at the 
production current of 10 nA. 
The magnetic
field of CLAS toroidal magnet was set to 1920 A, half of its maximum value, to optimize the momentum 
resolution and the efficiency for positively charged particles. 
A trigger was used with a required coincidence between hits in the tagger, the start
counter (ST), and the time-of-flight (TOF) paddles. 

\subsection{Event Selection}
\begin{figure}[htbp]
 \begin{center}
 \mbox{\epsfxsize=7.5cm\leavevmode \epsffile{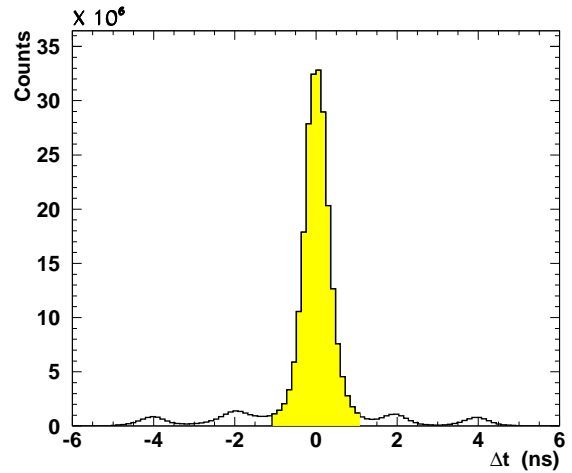}}	
 \end{center}
\caption{\small{(Color online) Difference between tagger and the start-counter (ST) times
(solid histogram). The tagger and ST coincidence time for the 
selected events 
is required to be within 1 ns (shaded histogram).
Secondary peaks corresponding to nearby beam buckets are also visible. }}
\label{fig:dt}
\end{figure}

In order to associate the reaction of interest with the triggering tagged photon,  
the coincidence time between the Tagger and CLAS was required to 
be within 1 ns. A cut was applied to the time  
difference, $\Delta t$, between the CLAS start time at the interaction point recorded by
the Start Counter (ST) and the Tagger.  
The central peak in Fig. \ref{fig:dt} corresponds to the tagger hits that are 
in time coincidence with CLAS within the 2-ns-wide beam bucket.   
In the g3a run period, only about 2$\%$ of
the events contained more than one tagged photon.

\begin{figure}[htbp]
 \begin{center}
 \mbox{\epsfxsize=8.5cm\leavevmode \epsffile{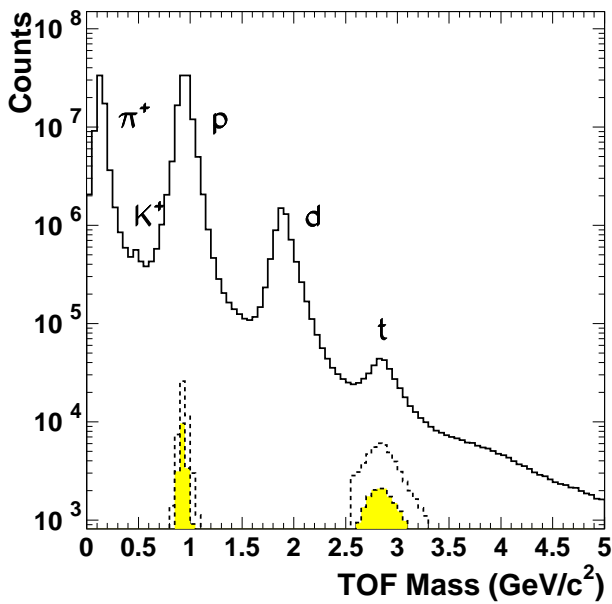}}	
 \end{center}
\caption{\small{(Color online) Hadron mass calculated from the time-of-flight 
and tracking information.
The solid histogram shows the mass distribution for all of the positively 
charged hadrons, the dashed histogram is the selected sample of protons and 
tritons that are detected in coincidence, and the shaded histogram 
shows the same distribution after applying all of the kinematic cuts to remove 
the background (see Sec. \ref{sec:background} for details). 
}}
\label{fig:tofmass}
\end{figure}

The particles were identified by determining their charge, momentum,
and velocity. Charge and momentum were obtained from the drift-chamber 
tracking information and the velocity from the time of flight
and path length to the time-of-flight detectors. Figure \ref{fig:tofmass}
shows the reconstructed mass distribution of positively charged particles. The 
events of interest were those with two and only two positively charged 
particles detected in coincidence. A triton candidate was required to
have a positive charge and a reconstructed mass squared $m^2$ between 6.5 
and 11.0 (GeV/{\it{c}}$^2$)$^2$. A proton candidate
was required to have a positive charge and a reconstructed mass
squared between 0.4 and 1.4 (GeV/{\it{c}}$^2$)$^2$.
In order to assure that the events of interest are produced within the
$^4$He target volume, a cut was applied to the $z$-component of the interaction 
vertex along the beam line.

\begin{figure}[htbp]
 \begin{center}
 \mbox{\epsfxsize=8.5cm\leavevmode \epsffile{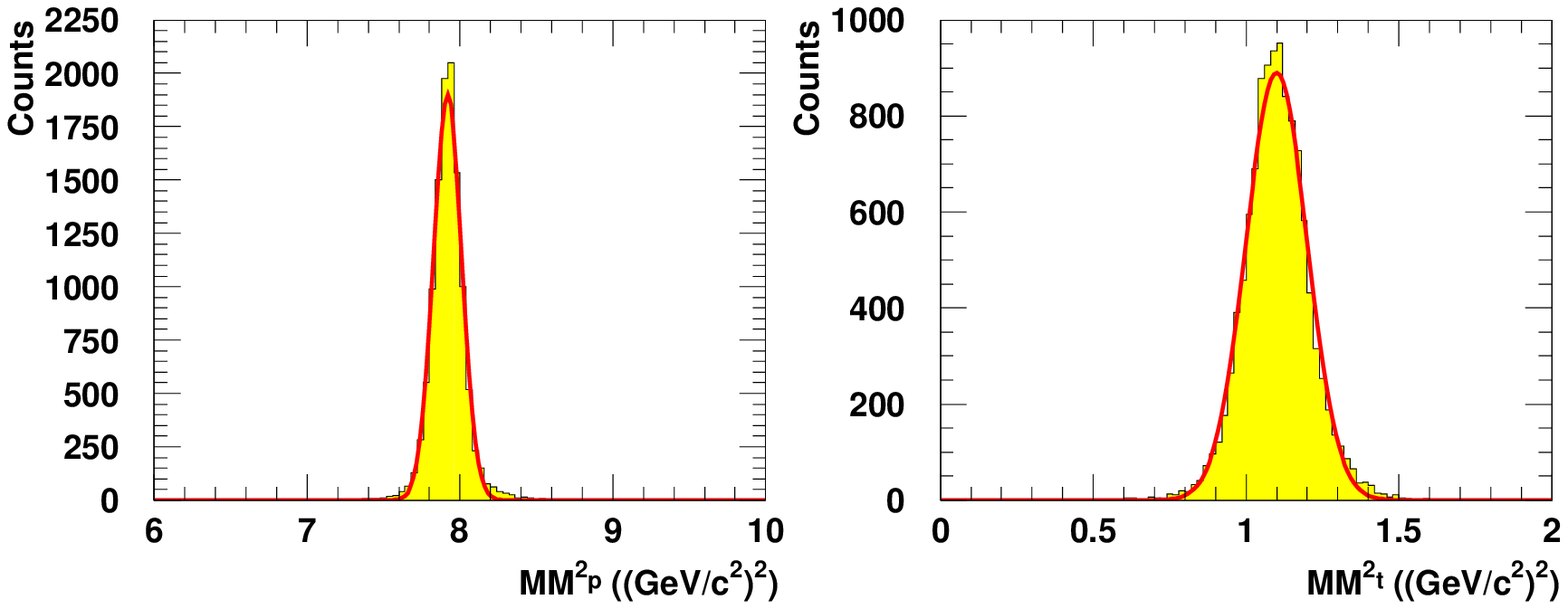}}
 \mbox{\epsfxsize=8.5cm\leavevmode \epsffile{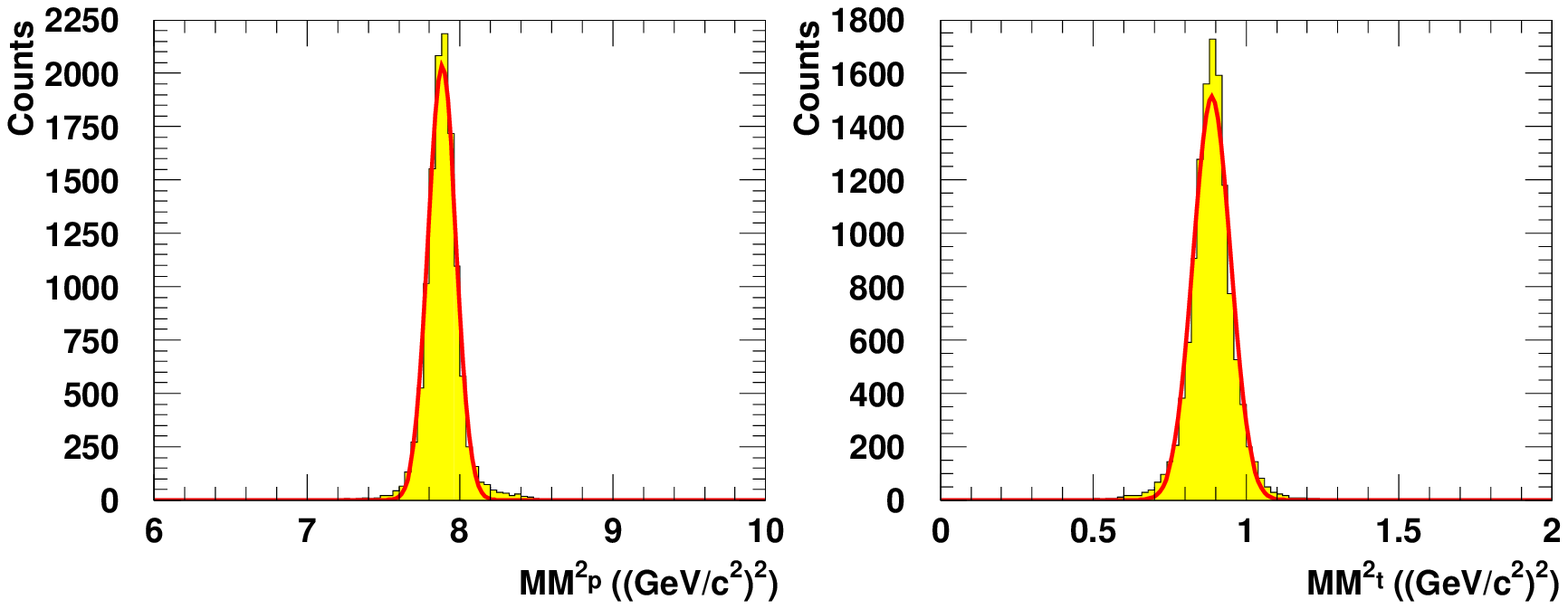}}	
 \end{center}
\caption{\small{(Color online) Distributions of the proton (left) and triton (right) missing-mass squared 
before (top) and after (bottom) the energy-loss corrections. Gaussian fits have been 
performed to determine the mean value and the width of each distribution (see Table \ref{tab:eloss}).}}
\label{fig:eloss}
\end{figure}

Energy-loss corrections were applied to the selected particles because
they lose a non-negligible part of their energy in the target material
and start counter before they reach the drift chambers.  
The effect of the energy-loss corrections after
applying all of the kinematic cuts on the final sample of $pt$ data 
is shown in Fig. \ref{fig:eloss}.  The importance  
of these corrections can be demonstrated by comparing the missing-mass squared of 
either the detected proton or the detected triton before and after applying 
these corrections. As expected, the amount of the energy loss for a particle 
depends on the mass of that particle and, therefore, these corrections have 
a larger effect on the measurement of the triton than of the proton.  
Table \ref{tab:eloss} summarizes the result of fitting Gaussians
to the  proton and
triton missing-mass-squared distributions before and after the energy-loss 
corrections.

Also, 
fiducial-volume cuts were applied to ensure that the particles are detected
within those parts of the volume of CLAS where the detection efficiency is high and uniform.
These cuts select regions of the CLAS where simulations reproduce
detector response reasonably well.

In order to eliminate any possible background, two-body kinematics were
used to select a clean sample of $pt$ events (see Sec. \ref{sec:background}).
.

\begin{table}[h]
\begin{center}
\caption[]{\small{Summary of the obtained mean values and widths of the proton and triton
missing-mass-squared distributions before and after the energy-loss corrections. The accepted
values for the proton and triton mass squared are 0.8804 and 7.890 (GeV/{\it{c}}$^2$)$^2$, respectively.}}
\label{tab:eloss}

\begin{tabular} {lcc} \hline \hline
		& without corrections & with corrections   \\ \hline
${MM^2}_p$ (GeV/$c^2)^2$	&   7.919 		& 7.883		\\ 
Width (GeV/$c^2)^2$  		&	0.09238		&	0.09127	 \\ \hline
${MM^2}_t$ (GeV/$c^2)^2$	&    1.100		&  0.8879		\\  
Width (GeV/$c^2)^2$ 		&	0.1001	&	0.06159	 \\ \hline \hline	
\end{tabular}

\end{center}
\end{table}
 
\subsection{Background Corrections}
\label{sec:background}
In order to select cleanly the $\gamma ^4$He $\rightarrow pt$ channel,
two-body kinematics were used. The two-body final-state kinematics 
for real events requires that the missing energy, missing momentum,
and missing-mass squared for $pt$ events be zero.  
Also, the opening angle between the three-vectors of the detected proton and triton
$\theta_{pt}$ should be close to 180$^{\circ}$ in the center-of-mass frame.
Our initial sample of events contains two and only two charged particles.
Four-vector conservation for the reaction $\gamma ^4$He $\rightarrow pt$,
as specified in Eq.(\ref{eq:fourvec}), leads to the determination of three 
kinematic variables --  
the missing energy $E_X$, the missing momentum
$P_X = \sqrt{P_X(x)^2 + P_X(y)^2 + P_X(z)^2}$, and the missing-mass squared
$M_X^2 = E_X^2 - P_X^2$: 
\begin{equation}
\left( \begin{array}{c}
{E_{\gamma}}  \\
0  \\
0 \\
{E_{\gamma}} 
\end{array} \right) + \left( \begin{array}{c}
{M_{^4He}}  \\
0  \\
0 \\
0 
\end{array} \right)  = 
\nonumber
\end{equation}

\begin{equation}
\left( \begin{array}{c}
\sqrt{m_p^2+p_p^2}  \\
p_p(x)  \\
p_p(y) \\
p_p(z) 
\end{array} \right) + \left( \begin{array}{c}
\sqrt{m_t^2+p_t^2}  \\
p_t(x)  \\
p_t(y) \\
p_t(z) 
\end{array} \right) + \left( \begin{array}{c}
E_X  \\
P_X(x)  \\
P_X(y) \\
P_X(z) 
\end{array} \right),
\label{eq:fourvec}
\end{equation}
where $E_\gamma$ is the incident photon energy, $M_{^4He}$ is the mass of the target nucleus,
$m_p$ and $m_t$ are the masses of the proton and triton, respectively, and $p_p$ and $p_t$ 
are the measured three-momenta of the proton and  triton, respectively.
These kinematic variables are plotted in 
Fig. \ref{fig:kinaftercuts}. For the real two-body break-up events into $pt$, we then have 
$E_X=0$ GeV, $P_X= 0$ GeV/{\it{c}}, $M_X^2 = 0$ (GeV/{\it{c}}$^2$)$^2$, and $\theta_{pt}= 180^{\circ}$.
Indeed, in Fig. \ref{fig:kinaftercuts} we see clear peaks showing the real two-body 
$pt$ break-up events.
However, some background can be seen in the selected events.
These events (mostly due to the $pt\pi^0$ channel) can be removed by applying additional 
kinematic cuts as follows:

\begin{figure}[htbp]
 \begin{center}
 \mbox{\epsfxsize=8.5cm\leavevmode \epsffile{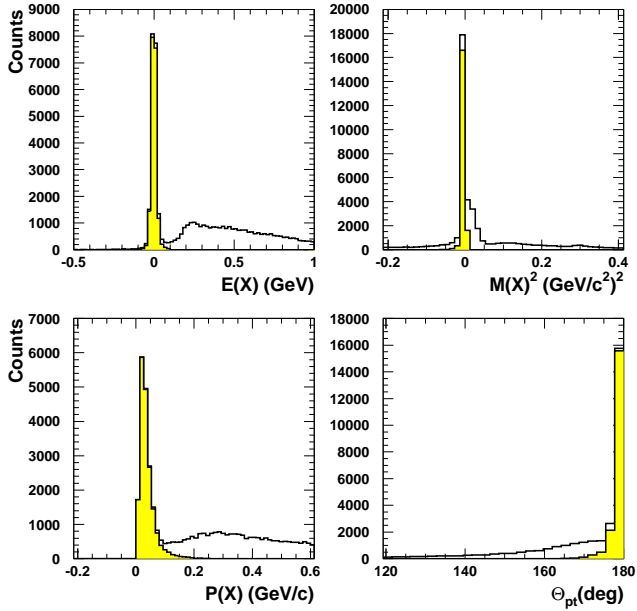}}	
 \end{center}
\caption{\small{(Color online) The $pt$ two-body final-state kinematics require the missing energy
(upper left), missing-mass squared (upper right), and missing momentum (lower left)
to be zero, and the $pt$ opening angle in the center-of-mass frame (lower right) to be 180$^{\circ}$. The peaks correspond
to the real $pt$ events from the two-body breakup of $^4$He. 
The shaded areas correspond to the nearly background-free sample of $pt$ events after 
the kinematic cuts described in this section were applied.
The upper left panel shows the cut applied to the missing energy for pt events; the
other three panels show the derived event distributions.}}
\label{fig:kinaftercuts}
\end{figure}

\begin{figure}[htbp]
 \begin{center}
 \mbox{\epsfxsize=8.5cm\leavevmode \epsffile{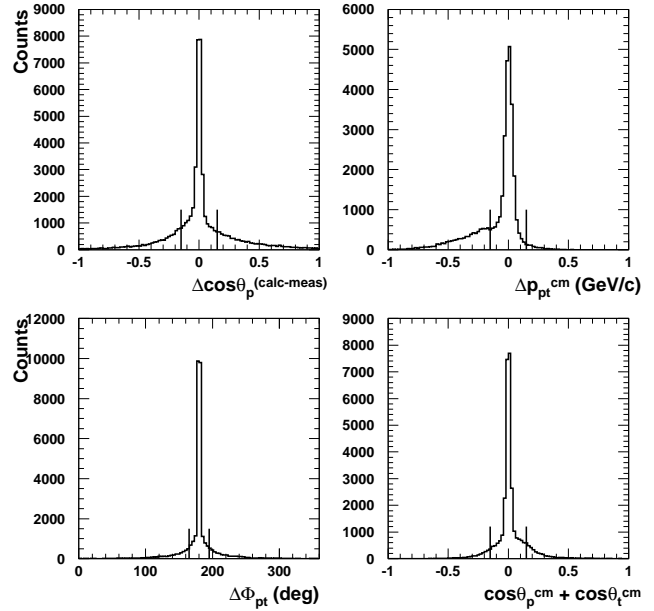}}	
 \end{center}
\caption{\small{Cuts were applied on various kinematic variables to remove
the background. Upper left: The difference between the measured and calculated
proton scattering angles. Upper right: The difference
between the magnitude of proton and triton momenta. Lower left: The difference between
the proton and triton azimuthal angles. Lower right: The sum of the cosines of the proton and 
triton scattering angles. All quantities are shown in the center-of-mass frame.}}
\label{fig:cuts}
\end{figure}

\begin{enumerate}
\item The first cut is applied to the difference between the measured scattering
angle of the proton in the center-of-mass frame (from the measured three-momentum 
vector of the proton) and the calculated one from the conservation of the four-momenta 
in the $\gamma ^4$He $\rightarrow pt$  reaction (by measuring only the 
triton momentum). This difference is plotted in the upper-left side of 
Fig. \ref{fig:cuts}. The clear peak around zero
corresponds to the real events from the two-body breakup of $^4$He into a proton and triton. 
The events for which this angular difference is outside of the range [-0.15,0.15] 
were removed from the data.

\item The second cut is applied to the difference between the momenta of the proton 
and the triton in the center-of-mass frame. For the real $pt$ events, this difference 
shows a peak around zero with a tail that could be due to the $pt\pi^0$ events,
as shown in the upper-right panel of Fig. \ref{fig:cuts}.
The applied cut requires this difference to be between -0.15 and 0.15 GeV/c.

\item The third cut requires the proton and triton three-momenta
to be in the same plane, \it{i.e.}, \rm the difference between the azimuthal angles for 
the proton and the triton in the center-of-mass frame is selected to be 
165$^{\circ}<\phi_{pt}^{cm}<$195$^{\circ}$. This distribution
is shown in the lower-left panel of Fig. \ref{fig:cuts}. A prominent peak around 180$^{\circ}$
is clearly seen.

\item The fourth cut is applied to the sum of the cosines of the proton and triton
scattering angles in the center-of-mass frame, shown in the lower-right panel of Fig. \ref{fig:cuts}. 
This cut retains only those events with -0.15$<$cos$\theta_p^{cm}+$cos$\theta_t^{cm}<$0.15.

\item Finally, the fifth cut requires the $pt$ missing energy to be 
-0.1$<E(X)<$0.1 GeV, shown in the upper left panel of Fig. \ref{fig:kinaftercuts}. 

\end{enumerate}

\begin{figure}[htbp]
 \begin{center}
 \mbox{\epsfxsize=9.0cm\leavevmode \epsffile{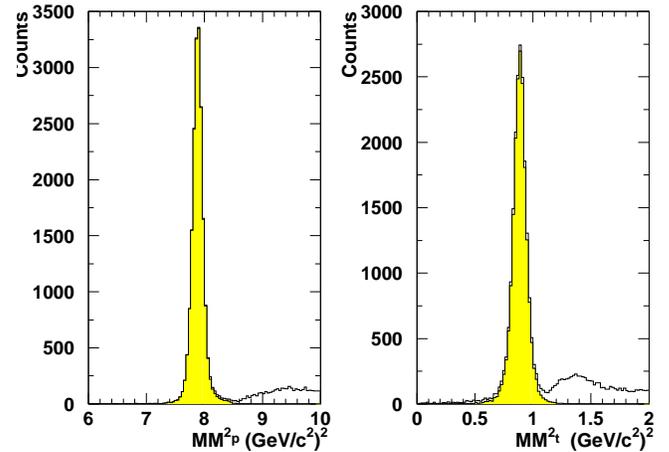}}	
 \end{center}
\caption{\small{(Color online) The calculated values for the missing-mass squared for the detected proton (left) 
and the detected triton
(right), before (solid histogram) and after (shaded histogram) applying the kinematic cuts. 
The background is completely removed by the kinematic cuts (see Sec. \ref{sec:background} for details).}}
\label{fig:mm}
\end{figure}

The value of each of these cuts is optimized such that the maximum number of ``good''
$pt$ events is retained. 
Using these cuts, the background in the spectra of the previously
described kinematic variables is mostly removed, as can be seen for 
the shaded areas of Fig. \ref{fig:kinaftercuts}.
The sample of events used after these cuts is therefore essentially background-free.
This also can be confirmed by calculating the missing-mass squared of either the detected 
proton or the detected triton. These distributions are shown before and 
after the above cuts in Fig. \ref{fig:mm}, and show that the background has been completely 
removed. The clean sample of protons and tritons that are detected in coincidence is also
shown within the shaded areas of Fig. \ref{fig:tofmass}. 

Table \ref{tab:cuts} summarizes the final cuts used to identify the
$pt$ events as described in this section.

\begin{table}[h]
\begin{center}
\caption[]{\small{Summary of kinematic cuts for event selection.}}
\label{tab:cuts}
\begin{tabular} {lc} \hline \hline
Description & Cut  \\ \hline
Coincidence time $\Delta t$	& $<$ 1 nsec   \\ 
Positively charged particles	& 2 			  \\ 
Proton identification		& $0.4<m_p^2<1.4$ (GeV/{\it{c}}$^2)^2$ \\ 
Triton identification   	& $6.5<m_t^2<11.0$ (GeV/{\it{c}}$^2)^2$ \\ 
z-vertex 			& [-8,8] (cm)		\\ 
$\Delta \cos\theta_p^{cm}$	& [-0.15,0.15]		\\ 
$\Delta p_{p,t}^{cm}$		& [-0.15,0.15] (GeV/{\it{c}})	\\ 
$\Delta \phi_{p,t}^{cm}$	& [165,195] deg	\\ 
$\cos\theta_p^{cm}+\cos\theta_t^{cm}$	& [-0.15,0.15]	\\ \hline \hline
\end{tabular}
\end{center}
\end{table}
\subsection{Detector Efficiency and Acceptance}

\begin{figure}[htbp]
 \begin{center}
 \mbox{\epsfxsize=8.5cm\leavevmode \epsffile{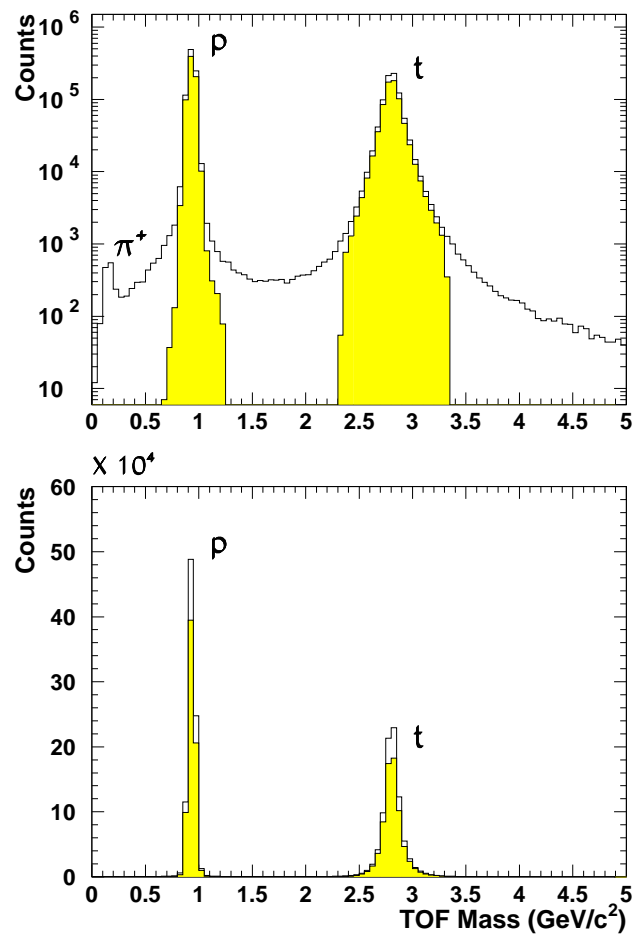}}
 \end{center}
\caption{\small{(Color online) Simulated TOF masses for Monte-Carlo generated
events, plotted with both logarithmic (top) and linear (bottom) scales, before (solid histogram) 
and after (shaded areas) applying all of the cuts.}}
\label{fig:tofmass-sim}
\end{figure}

The raw $pt$ yields are obtained as a function of the photon beam
energy $E_{\gamma}$ and the proton polar angle in the center-of-mass frame $\theta_p^{cm}$. 
The yields are corrected for the
detector acceptance using a Monte-Carlo simulation of phase-space-distributed
$pt$ events within the entire 4$\pi$ solid angle. The photon energy
was generated randomly with a uniform distribution from 0.35 to 1.55 GeV.
The standard GEANT-based CLAS simulation package \cite{gsim}) 
was used to simulate the detector response. 
The simulated events were processed with the same event-reconstruction
software that was used to reconstruct the real data. 
Figure \ref{fig:tofmass-sim} shows the reconstructed mass distributions for the 
simulated events with one proton and one triton after applying all of the cuts.

\begin{figure*}[htbp]
 \begin{center}
\hskip -1.0cm
 \mbox{\epsfxsize=14.0cm\leavevmode \epsffile{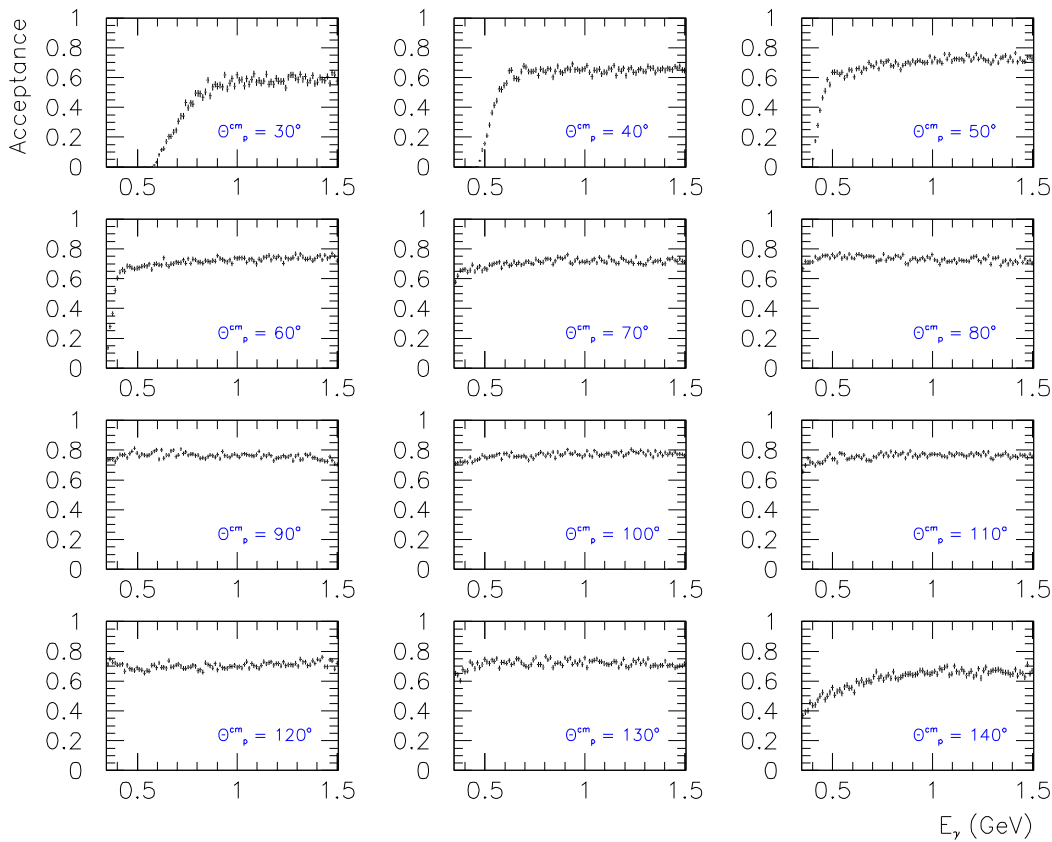}}	
 \end{center}
\caption{\small{(Color online) Acceptance as a function of photon
energy $E_\gamma$ for various proton-scattering-angle bins.}}
\label{fig:acc-fac-energy}
\end{figure*}

The acceptance is defined as the 
ratio of the number of reconstructed events to the number of generated
events.
Owing to the geometry and the structure of CLAS, there are regions of
solid angle that are not covered by the detector.
Furthermore, the inefficiencies in the various components of the detector
affect its acceptance and consequently the event reconstruction in CLAS. 
The acceptance correction factors are shown as functions of proton
scattering angle $\theta_p^{cm}$ and photon energy $E_{\gamma}$
for each kinematic bin in Fig. \ref{fig:acc-fac-energy}. 
These correction factors are used to 
convert the raw yields into unnormalized cross sections. 
Data points with poor acceptance ($<0.4$) at smaller angles 
are not included in the final data set.

\section{Results}

\subsection{Cross Sections}
The differential cross sections are obtained from the expression

\begin{equation}
{d\sigma \over d\Omega} = { N \over \eta_a N_{\gamma} N_T \Delta\Omega}
\end{equation}
where $N$ is the number of measured events in a given energy and angular bin of 
solid angle $\Delta\Omega = 2\pi\Delta \cos\theta_{cm}$. 
The CLAS acceptance is given by $\eta_a$; $N_{\gamma}$ is the number of 
photons within the given energy range incident on the target; and $N_T$
is the number of target nuclei per unit area. 

The number of target nuclei per unit area $N_T$ is determined from

\begin{equation}
N_T = {\rho l N_A \over A} \approx 3.0066 \times 10^{-10} \;\; \rm{nb}^{-1},
\end{equation}
where $l = 16.0$ cm is the target length, $\rho = 0.1249$ g/cm$^3$ is 
the density of liquid $^4$He, $A = 4.0026$ g/mole is its atomic
weight, and $N_A = 6.022 \times 10^{23}$ atoms/mole is Avogadro's
number. 

\begin{figure}[htbp]
 \begin{center}
\mbox{\epsfxsize=8.5cm\leavevmode \epsffile{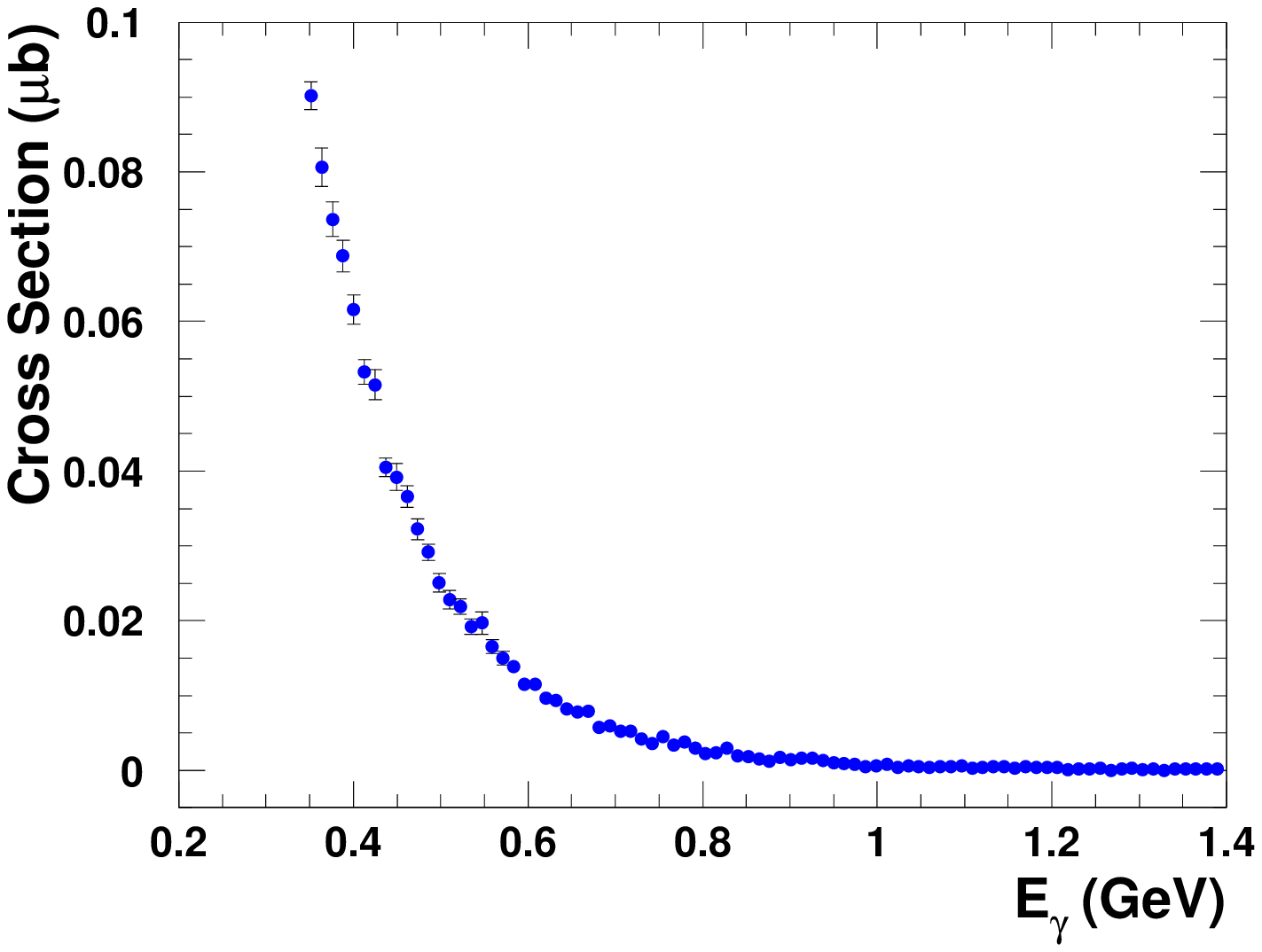}}
\mbox{\epsfxsize=8.5cm\leavevmode \epsffile{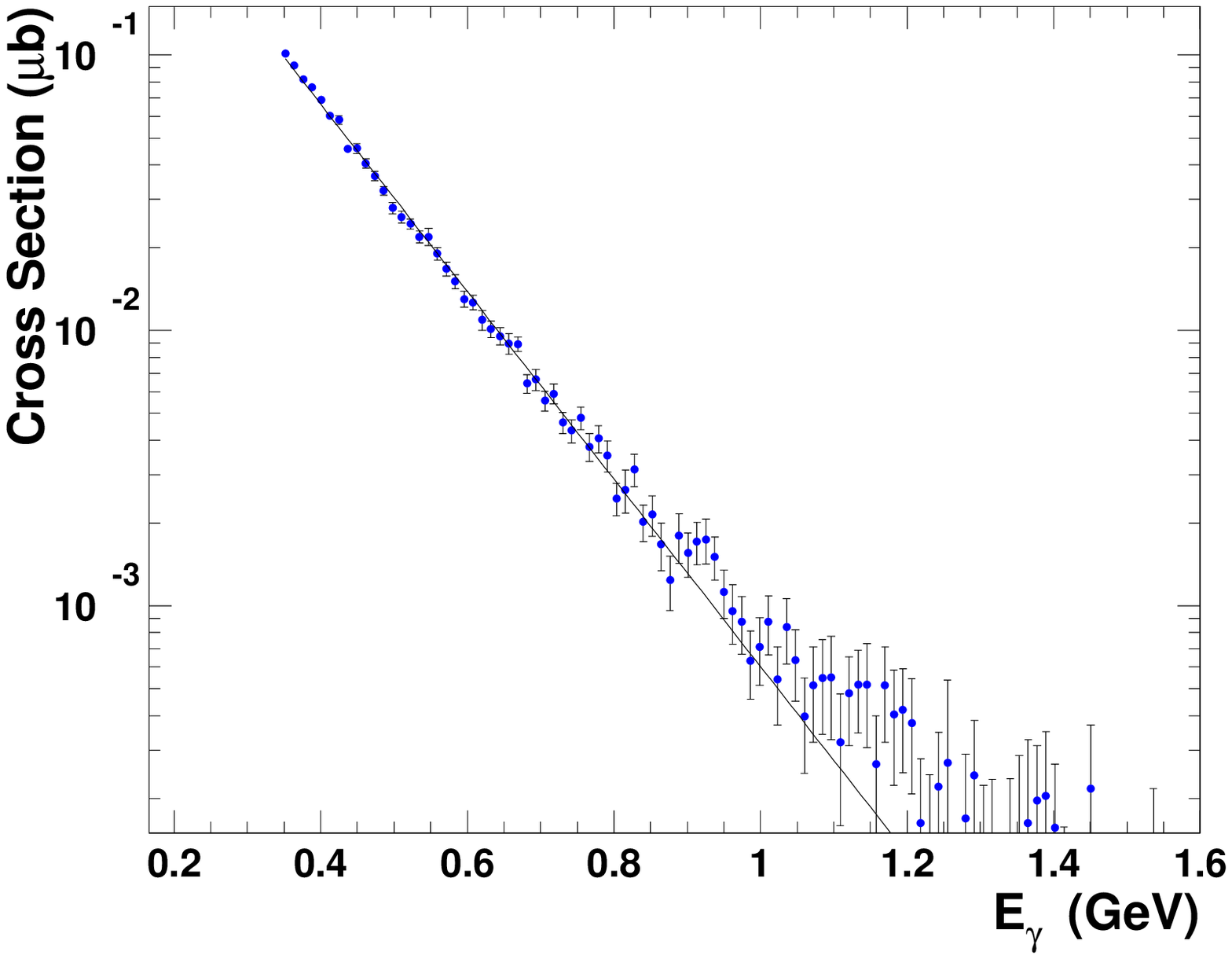}}
 \end{center}
\caption{\small{(Color online) Total angle-integrated cross section as 
a function of $E_\gamma$ in linear (top) and logarithmic (bottom) scales. 
Error bars indicate statistical uncertainties. The logarithmic plot was
fitted with an $Ae^{-BE_\gamma}$ functional form with $A=$1.35 $\mu$b, and $b=$7.8 GeV$^{-1}$.}} 
\label{fig:xsec-sum-energy}
\end{figure}

The photon yield $N_{\gamma}$ was obtained from the Tagger hits using the gflux 
analysis package
\cite{ball}. This number is corrected for the data-acquisition dead time. 
The angle-integrated cross section as a function of photon energy is shown 
in Fig. \ref{fig:xsec-sum-energy} in linear and logarithmic scales. The logarithmic
plot was fitted with an $Ae^{-BE_\gamma}$ functional form with $A=$1.35 $\mu$b, and $B=$7.8 GeV$^{-1}$.
It is remarkable that the total cross section follows an exponential
fall-off so closely, over the entire energy range from 0.4 to $\sim$1.0 GeV, 
flattening somewhat only above this energy, where forward angles dominate.

\begin{figure}[htbp]
 \begin{center}
\mbox{\epsfxsize=8.3cm\leavevmode \epsffile{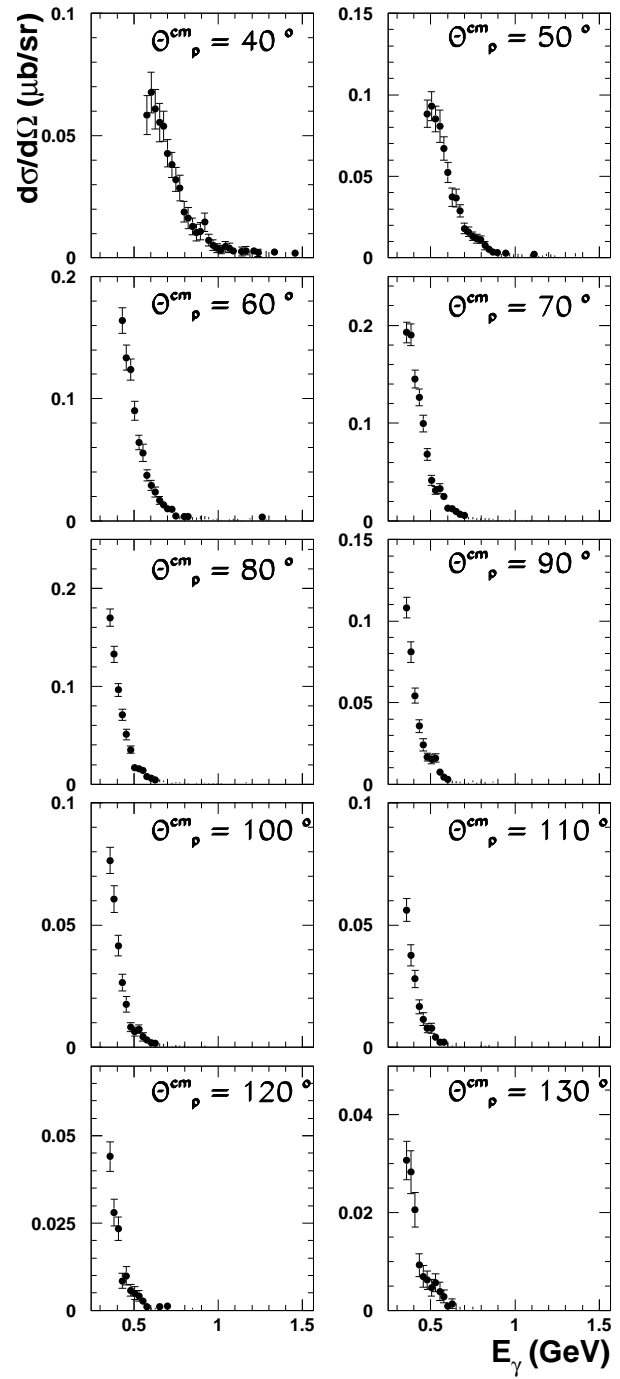}}
 \end{center}
\caption{\small{Measured differential cross sections as a function of $E_\gamma$ for
$\theta_p^{cm}$= 40, 50, 60, 70, 80, 90, 100, 110, 120, and 130 degrees.
Error bars indicate statistical uncertainties.}}
\label{fig:xsec-energy}
\end{figure}

\begin{figure}[htbp]
 \begin{center}
\mbox{\epsfxsize=8.3cm\leavevmode \epsffile{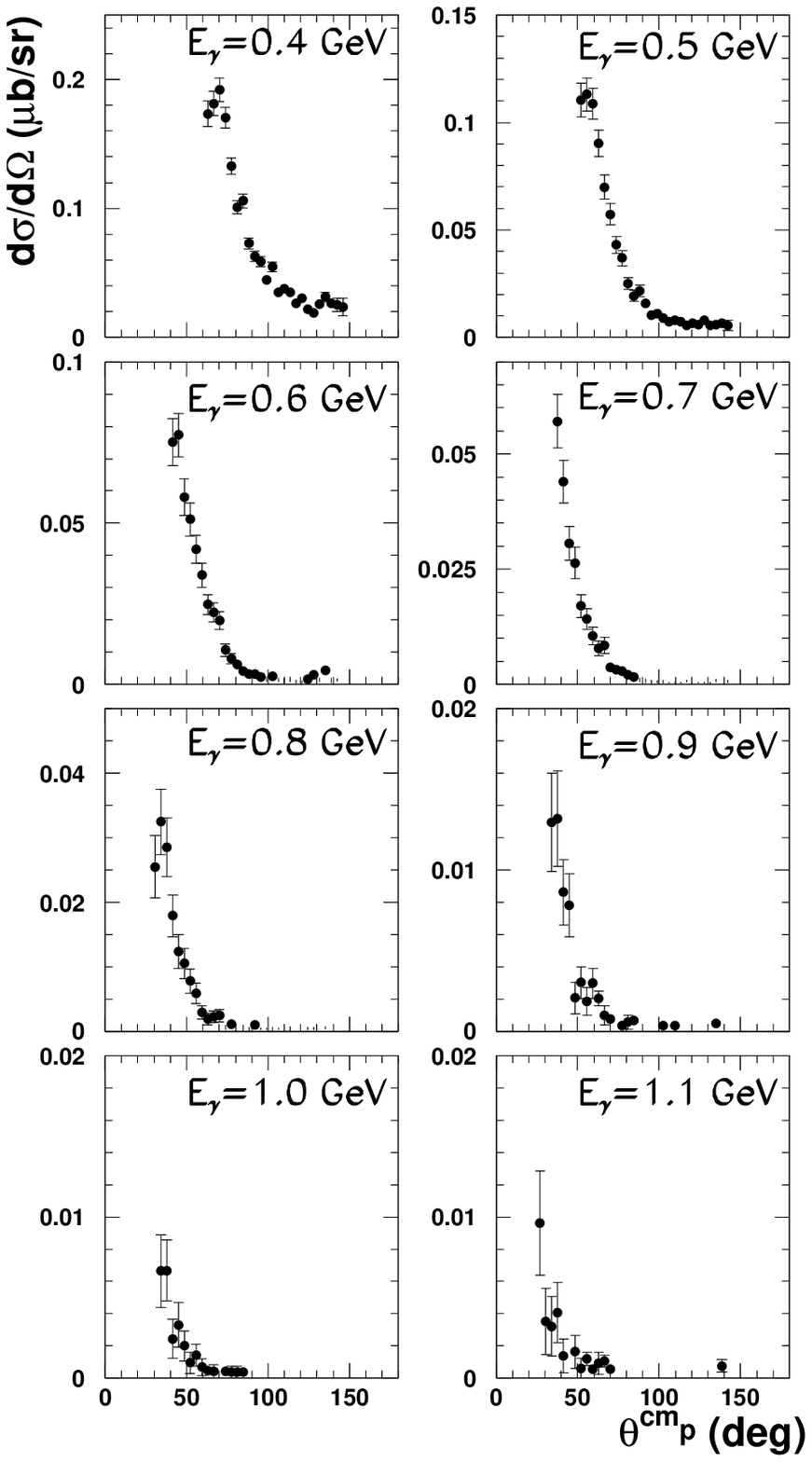}}
 \end{center}
\caption{\small{Measured differential cross sections as a function of 
$\theta_p^{cm}$ for $E_\gamma$ = 0.4, 0.5, 0.6, 0.7, 0.8, 0.9, 1.0, and 1.1 GeV.
Error bars indicate statistical uncertainties.}}
\label{fig:xsec-angle}
\end{figure}

The measured differential cross sections are shown in Figs. \ref{fig:xsec-energy} 
and \ref{fig:xsec-angle} as functions of photon energy and proton-scattering angles,
respectively.
These plots show that the peak of the angular
distributions shifts slightly towards smaller angles with increasing 
photon energy. 
\unboldmath


\boldmath
\subsection{Systematic Uncertainties}
\label{sec:syserr}
\unboldmath
Table \ref{tab:sys} summarizes the systematic uncertainties. The uncertainty in the 
photon-flux determination, including the tagger-efficiency evaluation,
is taken from the g3a analysis of Niccolai {\it et al.} \cite{niccolai}.
The value of the target density given in the literature was used; its uncertainty 
is no larger than 2$\%$. 
The uncertainties due to the 
fiducial cuts are estimated and have been found to be 
negligible.
\begin{table}[h]
\caption[]{\small{Summary of systematic uncertainties arising from various sources.}}
\label{tab:sys}
\begin{tabular} {lc} \hline \hline
Source			& Uncertainty ($\%$)    \\ \hline 
Photon flux		& 6 			\\ 
Target density		& $<$ 2 			\\ 
Solid angle 		& negligible 		\\ 
CLAS acceptance		& $<$ 10			\\ 
Fiducial cuts		& negligible		\\ 
Kinematic cuts 		& $<$ 10		\\    
\hline
Total			& $<$ 15			\\ \hline \hline			
\end{tabular}
\end{table}
The systematic uncertainty due to the CLAS acceptance was obtained
by comparing the cross sections measured by each pair of the
CLAS sectors independently (i.e., the data from sectors 1 and 4, 2 and 5,
and 3 and 6 were combined). The mean deviation between the three
sets of cross sections is considered to be an estimate of  the systematic
uncertainty for the CLAS acceptance.

In order to estimate the systematic uncertainty due to applying the
kinematic cuts, two sets of altered cuts, loose and tight, were used
and compared with the nominal cuts. The RMS (root mean square) of
the distribution of the differences between the cross sections obtained
with loose, tight, and the nominal cuts is considered to be a measure of the systematic
uncertainty due to these cuts.

The CLAS acceptance and kinematic cuts contribute the 
largest part of the systematic uncertainty.  
The individual systematic uncertainties are summed in quadrature to less 
than 15$\%$. The statistical uncertainties for the results usually dominate
the systematic uncertainties.

\begin{figure}[htbp]
 \begin{center}
 \mbox{\epsfxsize=8.7cm\leavevmode \epsffile{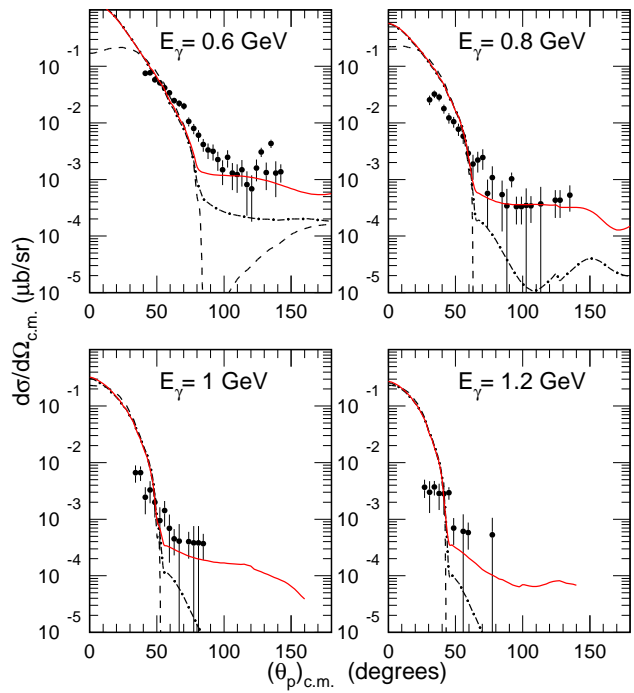}}	
 \end{center}
\caption{\small{(Color online) Measured differential cross sections compared with the 
Laget calculations \cite{laget2}.
The results are shown as a function of 
$\theta_p^{cm}$ for four photon-energy bins. 
The curves are calculations based on the
Laget model \cite{laget} including one-body (dashed line), two-body (dash dotted
line), and three-body (solid red line) mechanisms. Error bars indicate statistical 
uncertainties.
}}
\label{fig:xsec}
\end{figure}
\begin{figure}[htbp]
 \begin{center}
\vskip 0.5cm
 \mbox{\epsfxsize=8.0cm\leavevmode \epsffile{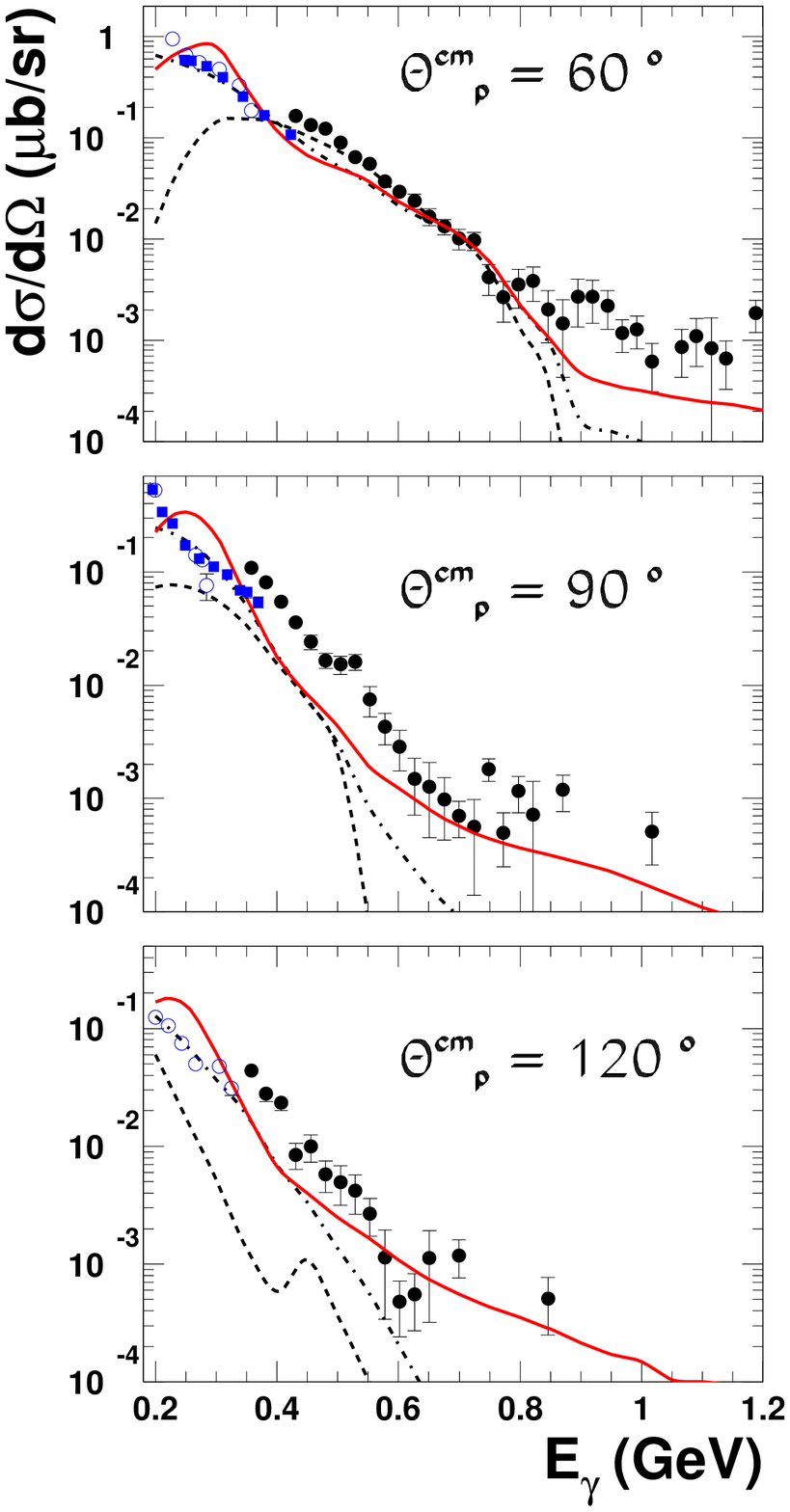}}	
 \end{center}
\caption{\small{(Color online) Measured differential cross sections (closed circles)
compared with the data from Ref. \cite{argan} (closed squares),  
Ref. \cite{schumacher} (open blue circles), and 
the Laget calculations \cite{laget2}, as a function of photon energy for three 
proton center-of-mass angles. The curves are as described in 
Fig. \ref{fig:xsec}. Error bars indicate statistical 
uncertainties. 
}}
\label{fig:xsec-data}
\end{figure}
\section{Discussion and Summary}
\label{sec:result}
Our bin-averaged cross sections are compared in Figs. \ref{fig:xsec} and \ref{fig:xsec-data} 
with the latest model calculations
by J.-M. Laget \cite{laget2}. Figure \ref{fig:xsec} shows the cross sections as a 
function of proton-scattering angle in the center-of-mass frame for four 
energy bins, centered at 0.6, 0.8, 1.0, and 1.2 GeV. Each energy bin is 100 MeV
wide.

In general, for all photon-energy bins, our data are in qualitative
agreement only with the calculations that include three-body mechanisms,
confirming the importance of the contribution
of three-body mechanisms at these energies. At 0.6 and 0.8 GeV, our data are higher
than the calculations between 60 and 100 degrees, and between 60 and 80 degrees, 
respectively.
Our data are
consistent with the three-body calculations at large
angles for all photon-energy bins. 
However, at the higher of these energies, and at small proton
angles, there are discrepancies between the data and the results of the calculations.
This disagreement at the most forward angles at higher
energies might result from either of two factors: 
1) Off-mass-shell
effects might become significant when the energy increases.
Although the elementary electromagnetic operator (the coupling
of the photon to the nucleon) used in this model is the fully 
relativistic version of the coupling of the photon to an on-mass-shell
nucleon, the actual photon energy that enters the amplitude is 
not the same as the photon energy when the target proton is at
rest. The difference between the two energies increases with
the photon energy. This is a well identified problem that still
has no definite solution. 2) The previous drawback is less
severe for the three-body amplitude that dominates at large angles.
Here, the momentum transfer is shared between the three nucleons and 
one probes mainly the low-momentum components
of the $^4$He wave function. Also, the amplitude depends mainly on the
elementary processes, where the nucleons are mostly on-shell.

The cross sections are shown in Fig. \ref{fig:xsec-data}
as a function of photon energy for three angular bins
centered at 60, 90, and 120 degrees. Each angular bin is 10 degrees wide.
Comparison with Fig. \ref{fig:laget}
shows that 
the trend of the data and calculations
are very similar. However there is a clear difference between the relative
strength of the two- and three-body contributions with respect
to the one-body contribution shown in the two plots. 
This is because the earlier version of the Laget model \cite{laget} 
used the AV14 potential \cite{av14}. This version included fewer high-momentum 
components than the newer version, which uses AV18 \cite{av18}. Comparison 
shows that in general our data stand above
all the curves especially at 90 and 120 degrees
but qualitatively are in good agreement with the calculations. 

Figure \ref{fig:xsec-data} also shows some of the earlier experimental data
from the Saclay group \cite{argan} and the MIT group \cite{arends}, compared with the 
results of this experiment.
There is a very limited overlap in the photon energy range
between our data and the older data from Saclay.  
The range of the overlap is from 423 to 430 MeV for 60 
degrees, and from 357 to 369 for 90 degrees. 
There is no overlap at all between our data and older data from MIT.
The comparison shows a continuous trend with increasing photon energy
for the previous data to lie below our data in the overlap region.

In summary, we have measured the differential cross sections for 
the $\gamma ^4$He $\rightarrow pt$ reaction
in the energy range from 0.35 to 1.55 GeV, for proton center-of-mass 
scattering angles between 40 and 140 degrees. 
It is important to emphasize that the interpretation of these data  
is model dependent. We have compared them with the results of the only 
available theoretical calculation at these energies \cite{laget2}. 
This comparison reveals the essential importance of the contribution of 
three-body mechanisms, especially in the energy region of 0.6-0.8 GeV, as
was found previously for $^3$He \cite{niccolai}. 
These data are important
for understanding the reaction mechanisms and for developing models of this
process for photon energies above 0.4 GeV, and even more important for an understanding 
and appreciation of the importance, in the relevant range of energy and wavelength, 
of strong many-body forces in nuclei.


\section*{Acknowledgments}

We would like to acknowledge the outstanding efforts of the staff of 
the Accelerator and the Physics Divisions at Jefferson Lab that made 
this experiment possible.  This work was supported by the U.S.
Department of Energy under grant DE-FG02-95ER40901.
the National Science Foundation, the Italian 
Istituto Nazionale di Fisica Nucleare, the French Centre National de la 
Recherche Scientifique, the French Commissariat \`{a} l'Energie 
Atomique, and the Korean Science and Engineering Foundation, and  
the UK Science and Technology Facilities Council (STFC).
The Southeastern Universities Research Association (SURA) operated the 
Thomas Jefferson National Accelerator Facility for the United States 
Department of Energy under contract DE-AC05-84ER40150. 


\end{document}